# Why Nanosystems and Macroscopic Systems Behave Differently

Pirooz Mohazzabi
Department of Physics
University of Wisconsin-Parkside
Kenosha, Wisconsin 53141, USA
pirooz.mohazzabi@uwp.edu

and

G. Ali Mansoori
Departments of Bio & Chemical Engineering
University of Illinois at Chicago
(M/C 063) Chicago, Illinois 60607-7052, USA
mansoori@uic.edu

**Abstract**

Advancement in nanotechnology is dependent on the understanding of the behavior of matter in nanoscale. In this report we have demonstrated, through a unique molecular simulation procedure, that properties of matter in nanosystems do not follow the same rules as in macroscopic systems. More specifically, it is shown that extensive properties such as internal energy and entropy are not extensive, and intensive properties such as temperature and pressure are not intensive in nanosystems, in contrast to macroscopic systems. Variations of the Euler exponents for these properties as a function of the number of particles of the system are reported. As the size of the system increases, the value of these exponents approach integers that correspond to the macroscopic systems.

---

**Introduction:**

On December 29, 1959, Richard Feynman gave a classic talk entitled "There's Plenty of Room at the Bottom" at the annual meeting of the American Physical Society at the California Institute of Technology [1]. The idea that was truly ingenious set the landmark and opened the doors to a new era in science and technology at the nanoscale. Ever since and especially towards the end of the 20th century, nanoscience has grown explosively due to its tremendous potential applications in various areas, ranging from medicine to space technology.

In recent years hundreds and perhaps thousands of articles have appeared in a variety of highly regarded national and international media, reporting experimental, theoretical, and computational investigations related to the science at the nanoscale. For an example see [2,3].

The thermodynamics and statistical mechanics of macroscopic systems are well understood and have been known for a long time. That for small systems, on the other hand, was first introduced by Hill in 1963 and 1964 [4,5], which is a brilliant attempt of relating macroscopic and small systems. However, for small systems, the validity of the relationships among various thermodynamic quantities that apply to macroscopic systems is an open question.

One of the fundamental concepts in the thermodynamics of macroscopic systems is the number of independent variables of the system, which is the smallest number of properties that must be known in order to completely specify the thermodynamic state of the system. Among other differences, in small systems the number of independent variables is not the same as that for the corresponding macroscopic system.

Small systems of interest in nanotechnology are generally made up of condensed matter of various kinds, soft or hard, organic or inorganic, and/or biological components. They are normally isolated nanostructures and their assemblies, small droplets and bubbles, clusters, and fluids and solids confined in small regions of space like inside nanotubes and fullerene. Nanosystems are larger than individual molecules but smaller than micro systems. One of the main characteristics of nanosystems is their high surface-to-volume ratio. Furthermore, their electronic and magnetic properties are often distinguished by quantum behavior and their mechanical and thermal properties can be formulated within the framework of classical statistical mechanics of small systems [4-9] and through the newly developed field of nonextensive statistical mechanics [9,10].

When two macroscopic systems in equilibrium are combined, some of the properties double and some remain the same. For example, the volume and the internal energy of the combined system are twice those of the individual systems. Temperature and pressure of the combined system, on the other hand, are the same as those of the individual systems. The former thermodynamic properties are called *extensive* whereas the latter are referred to as *intensive*. Thus, if two identical macroscopic systems in equilibrium, each with an extensive parameter $X_1$ and an intensive parameter $I_1$, are combined to form a new system twice as large, the extensive and intensive parameters of the new system will be $X_2 = 2X_1$ and $I_2 = I_1$, respectively. In other words, for a general parameter $Z$, we can write $Z_2 = 2^{\lambda} Z_1$, where $\lambda = 1$ if $Z$ is an extensive parameter and $\lambda$

= 0 if $Z$ is an intensive parameter. We shall refer to $\lambda$ as Euler exponent as it is defined through Euler's theorem of homogeneous functions [11].

In a recent paper Vakili-Nezhaad [12] has proposed that the Euler's homogeneous functions with a non-integer order can define the thermodynamic functions of the nonextensive systems. Non-extensive thermodynamics, which is used by Vakili-Nezhaad, is according to the statistics introduced by Tsallis [9] in 1988. In the present report we prove the nonextensivity of small systems in the framework of the well-known Boltzmann-Gibbs statistics and through a molecular dynamics simulation.

## Theory

In what follows, we describe the results of an investigation that shows in nanosystems the notion of extensivity and intensivity breaks down; macroscopic extensive parameters become nonextensive and macroscopic intensive parameters become nonintensive in small systems. In fact, we shall see that thermodynamic parameters in small systems normally have Euler exponents that are nonintegers, or even negative.

We investigated three-dimensional microcanonical ensembles of systems of identical argon-like particles. The particles are confined to a fixed volume $V$ and interact elastically according to the pairwise Lennard-Jones (6-12) interatomic potential energy function. Collisions of the particles with the walls of the container are also elastic, therefore, the internal energy of the system is a constant. We investigated systems containing $n^3$ particles (where $n = 2,3,4, \ldots,10$), first individually, and then by combining identical such systems in pairs [13].

The temperature of the system is calculated by taking the time average of the mean kinetic energy per particle of the system [14]. The pressure is evaluated from the direct definition of pressure, i.e., change of the momentum of the particles per unit time per unit area as a result of their collisions with the walls of the container. The entropy of the system is computed using the general definition of entropy [15],

$$S = -k\sum_i p_i \ln p_i \qquad (1)$$

where $p_i$ is the probability of the microstate $i$, and the summation is over all possible microstates compatible with the system.

Figures 1-4 show the results of the simulations. In each case, two identical systems of $n^3$ particles each ($n = 2,3,4, \ldots,10$), are combined and the thermodynamic parameters of the new system are compared with those of the component systems. The Euler exponents are then calculated from

$$\lambda = (\ln 2)^{-1} \ln\left(\frac{Z_2}{Z_1}\right) \qquad (2)$$

Figure 1 shows the Euler exponent for internal energy when two exactly identical cubic systems are combined, by placing them together side by side, as a function of particle number. The nonextensivity is relatively large for small systems but decreases and approaches zero as the size of the component systems increase. Nonextensivity of the internal energy in small systems is a consequence of surface effects, which can be explained using nearest-neighbor interactions as follows: Consider a three-dimensional system of $N$ particles in the form of a cube. The particles are initially on a simple-cubic lattice with nearest-neighbor interactions among them. Out of $N$ particles, there are $6N^{2/3}$ particles on the faces of the cube, $12N^{1/3}$ particles on the edges of the cube, and 8 particles on the corners. However, since some of the particles in each category are shared between more than one region, these numbers are reduced by some fraction in each case. Thus, there are $6\alpha N^{2/3}$, $12\beta N^{1/3}$, and $8\gamma$ particles, respectively, on the faces, edges, and corners of the cubic system, where $\alpha$, $\beta$, and $\gamma$ are each a positive number less than one but greater than zero. Therefore, the number of bulk particles would be $N - 6\alpha N^{2/3} - 12\beta N^{1/3} - 8\gamma$.

If the interaction potential energy between each pair of neighboring particles is $-\varepsilon$, then the total internal energy of the system is given by

$$E = K - \frac{\varepsilon}{2}[6(N - 6\alpha N^{2/3} - 12\beta N^{1/3} - 8\gamma) + 5(6\alpha N^{2/3}) + 4(12\beta N^{1/3}) + 3(8\gamma)] \tag{3}$$

where $K$ is the total kinetic energy of the particles, and the factor 1/2 in front of the second term is introduced to avoid counting each interaction twice. Equation (3) simplifies to

$$E = K - 3N\varepsilon(1 - \alpha N^{-1/3} - 4\beta N^{-2/3} - 4\gamma N^{-1}) \tag{4}$$

Suppose that we have two exactly identical but isolated (infinitely apart) such systems. The energy of each system is given by Eq. (4). If we bring the two systems together, the total internal energy of the combined system is given by $E_2 = 2E_1 + \delta E$, where $E_2$ is the internal energy of the combined system and $E_1$ is that of each component. The additional energy, $\delta E$, comes from the interaction of the particles on near faces of the two systems that are placed together, as shown schematically in Figure 5. More specifically, some *external* potential energy is converted into internal energy. This energy is $\delta E = -N^{2/3}\varepsilon$. Therefore,

$$E_2 = 2K - 6N\varepsilon\left[1 + (\frac{1}{6} - \alpha)N^{-1/3} - 4\beta N^{-2/3} - 4\gamma N^{-1}\right] \tag{5}$$

As we can see, $E_2 \neq 2E_1$, and the system is nonextensive with respect to the total energy. The term

$$(-6N\varepsilon)(\frac{1}{6}N^{-1/3}) = -\varepsilon N^{2/3} \tag{6}$$

on the right hand side of Eq. (5), which is a negative quantity, is the contribution to nonextensivity.

If we now increase the size of the system by increasing the number of particles of the system without bound, we can see from Eq. (4) that

$$E_1(\infty) = \lim_{N\to\infty} E_1 = K - 3N\varepsilon \tag{7}$$

And from Eq. (5) we see that

$$E_2(\infty) = \lim_{N\to\infty} E_2 = 2K - 6N\varepsilon = 2E_1(\infty) \tag{8}$$

This simple analysis shows that a small system is nonextensive with respect to internal energy but as the size of the system increases, it eventually becomes extensive in the macroscopic limit.

Figure 3 shows the Euler exponent for temperature. Again the nonextensivity is relatively large for small systems but decreases as the number of particles increases. More specifically, for two identical systems each of 8 particles joining together, the temperature of the combined system is about 79% higher than the temperature of the constituent components. The difference decreases to about 2% when the number of particles in each system increases to 1000.

The increase in the temperature of the system can be explained as follows: when the two systems are brought together, some of the external potential energy between the two systems is transformed into the internal energy, making it more negative as we have seen in the previous section. This decrease of the internal energy increases the kinetic energy of the particles, thus increasing the temperature of the system.

Figure 4 shows the variation of the Euler exponent for pressure with the size of the system. Again, for small systems combining, the pressure of the new system is higher that that of the component systems before the combination. The increase in the pressure is attributed to the increase in the temperature of the system. At a higher temperature, the average speed and hence the average linear momentum of the particles is higher. Since pressure is the time rate of change of linear momentum of the particles per unit area as they collide with the walls of the container, it must increase when temperature increases.

The pressure of the systems, however, exhibits a very interesting effect: The Euler exponent decreases as the number of particles increases and becomes *negative* before approaching zero. This behavior is not a consequence of statistical fluctuation of the data. Indeed, for 729 particles, the value of the Euler exponent for pressure is -0.095 ± 0.002. No conclusion is intended to be drawn at this point regarding this anomaly.

Figure 2 shows the entropy which, like the internal energy, is nonextensive in small systems. There are, however, two differences between them. First, the nonextensivity in the entropy is not as profound as in the internal energy. For two systems of 8 particles each combining, the nonextensive contribution to the entropy is only about 4.4% as compared to 22% for internal energy. Similarly, when the size of each system increases to 1000 particles, the nonextensivity of the entropy drops to about 0.10% as compared to 3.0% for internal energy. Second, while the internal energy decreases (becomes more negative) due to nonextensivity [10], the entropy increases.

The superextensivity of entropy in small systems can be explained as follows: Because in small systems the temperature of the combined system is somewhat higher than that of the individual systems, the velocity distribution functions widen and shorten slightly. This in turn causes the entropy of the system to increase slightly above and beyond the sum of the entropies of the individual components. At first, this phenomenon seems to violate the second law of thermodynamics. It is, however, imperative to realize that conversion of some of the external potential energy associated with the separation of the two component systems into their internal energy is responsible for this behavior. As pointed out earlier, this additional energy is given by $\delta E = -N^{2/3}\varepsilon$.

## Conclusion

In conclusion, in small systems deviations from extensivity and intensivity of thermodynamic parameters are primarily domino effects that are initially triggered by conversion of some of the external potential energy into the internal potential energy of the combining systems, an effect that is insignificant in macroscopic systems. Furthermore, traces of nonextensivity and nonintensivity in thermodynamic parameters persist even in systems with as many as 1000 particles.

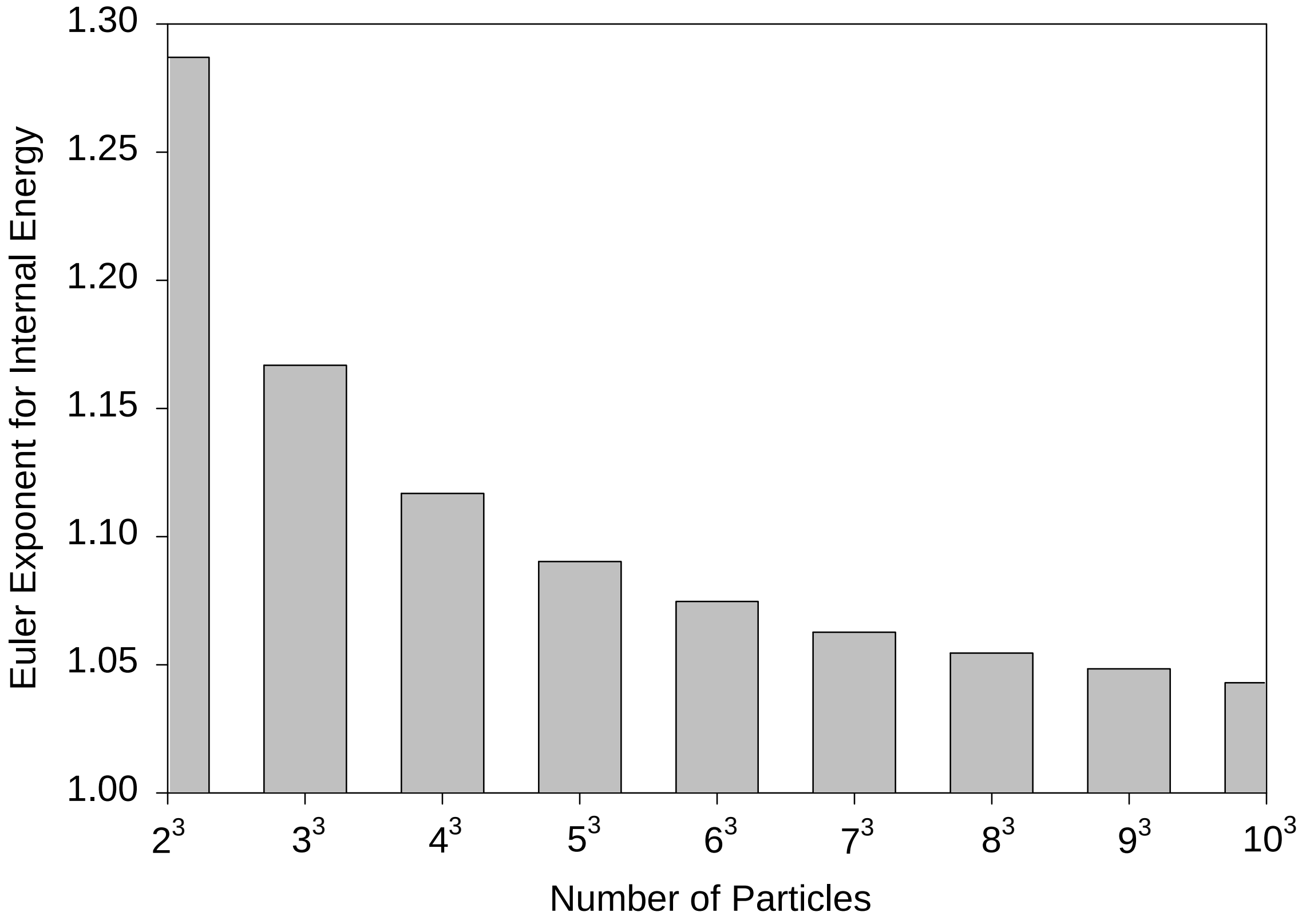

**Figure 1** - Euler exponent for internal energy as a function of system size.

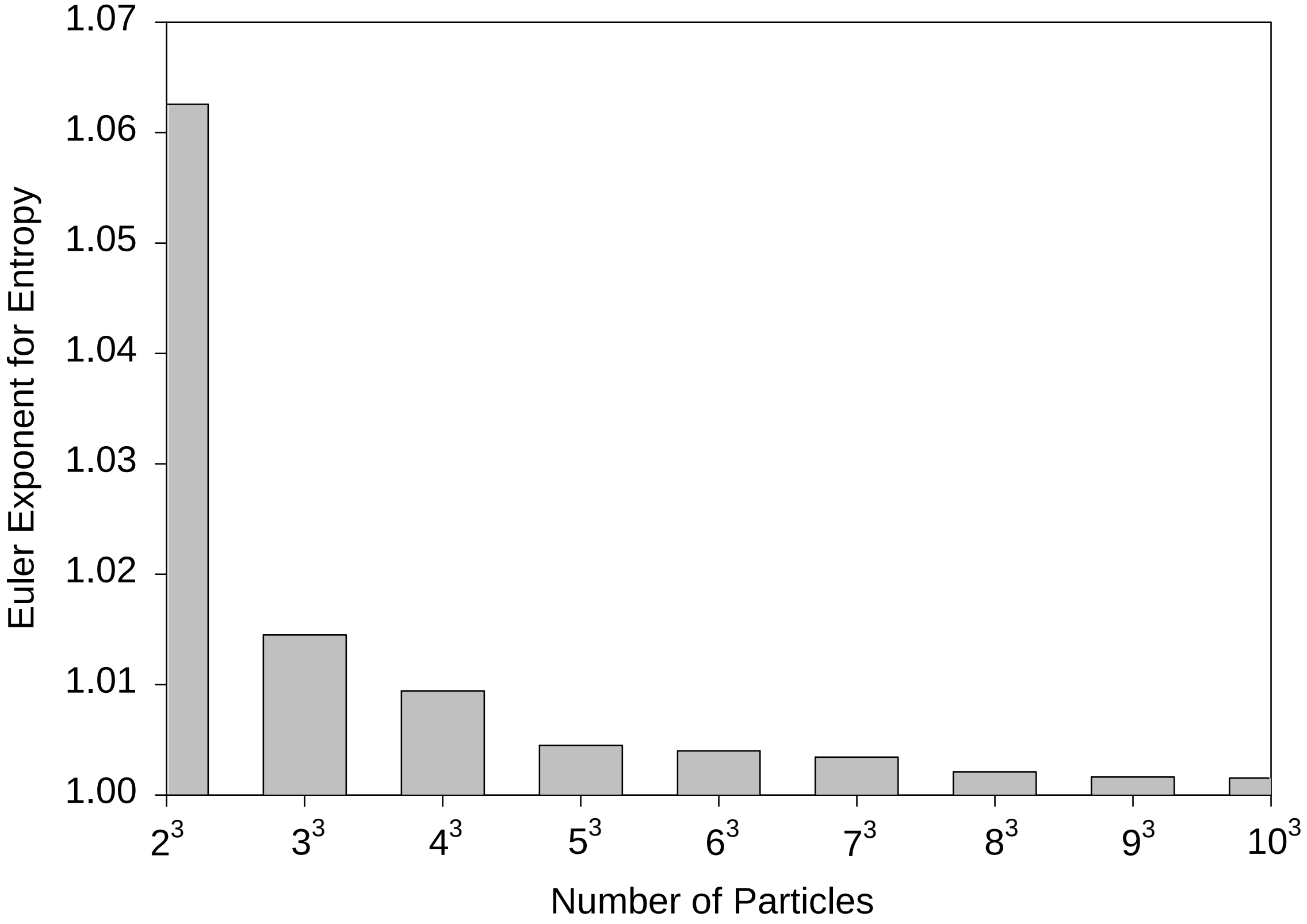

**Figure 2** - Euler exponent for entropy as a function of system size.

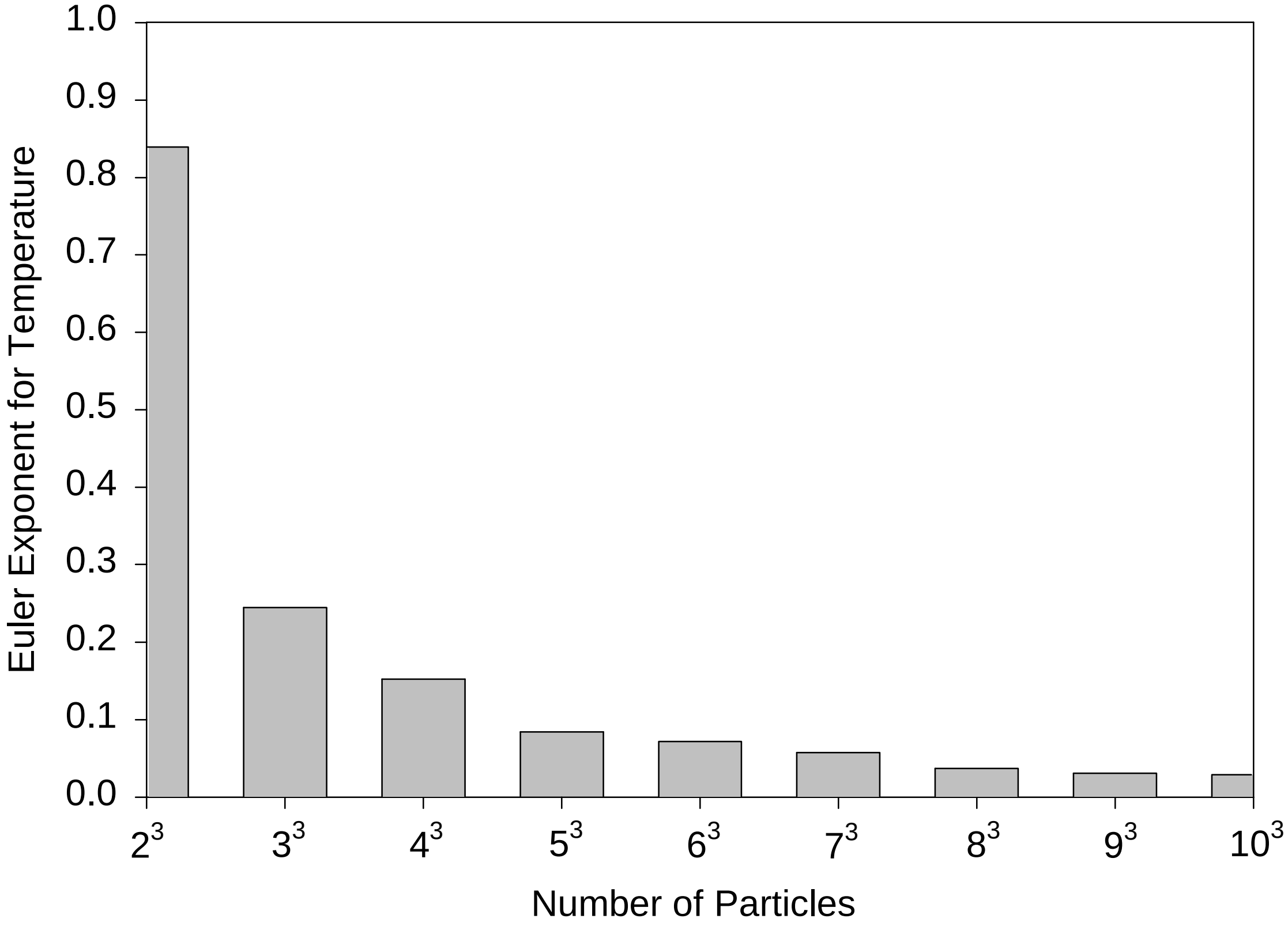

**Figure 3** - Euler exponent for temperature as a function of system size.

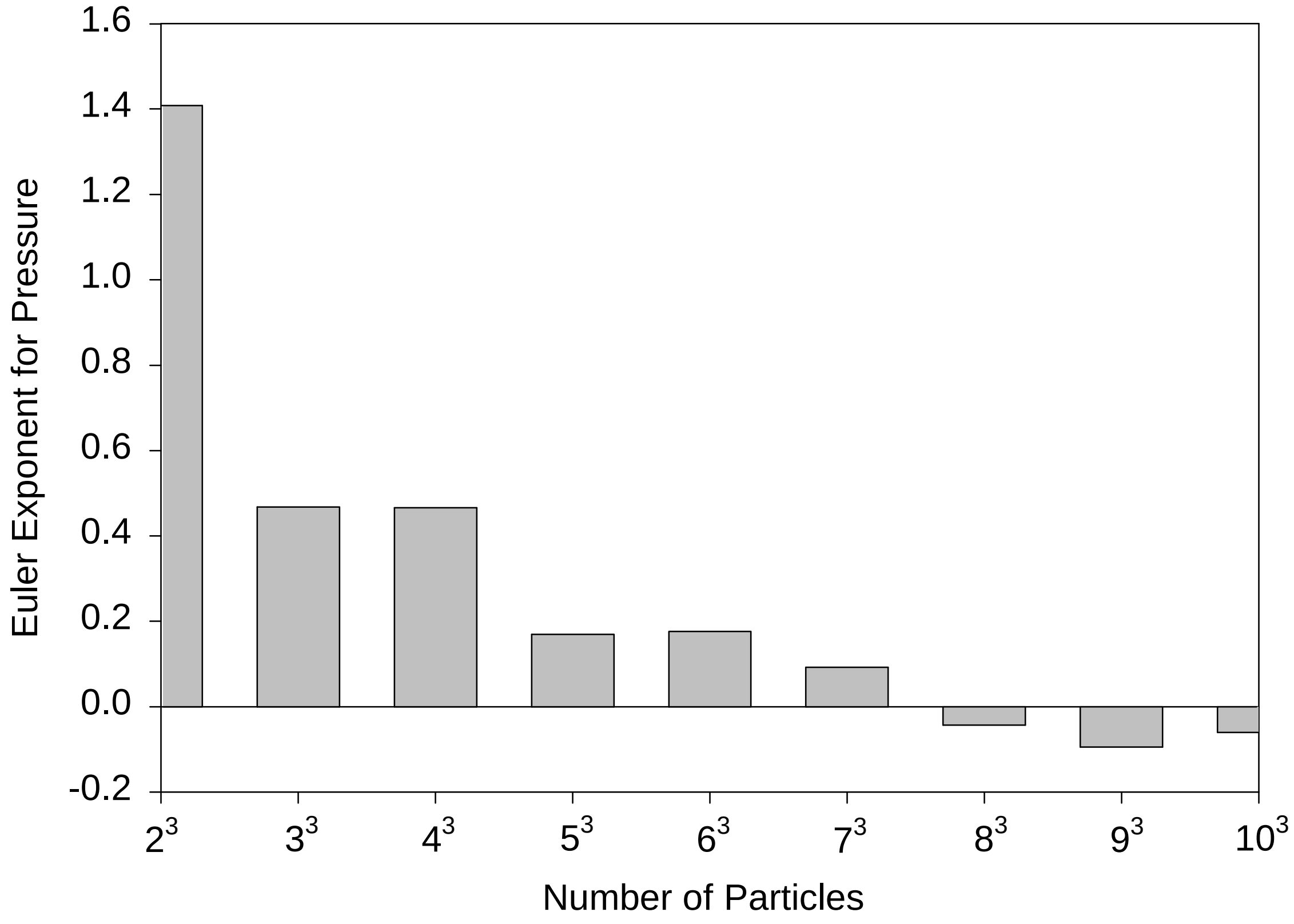

**Figure 4** - Euler exponent for pressure as a function of system size.

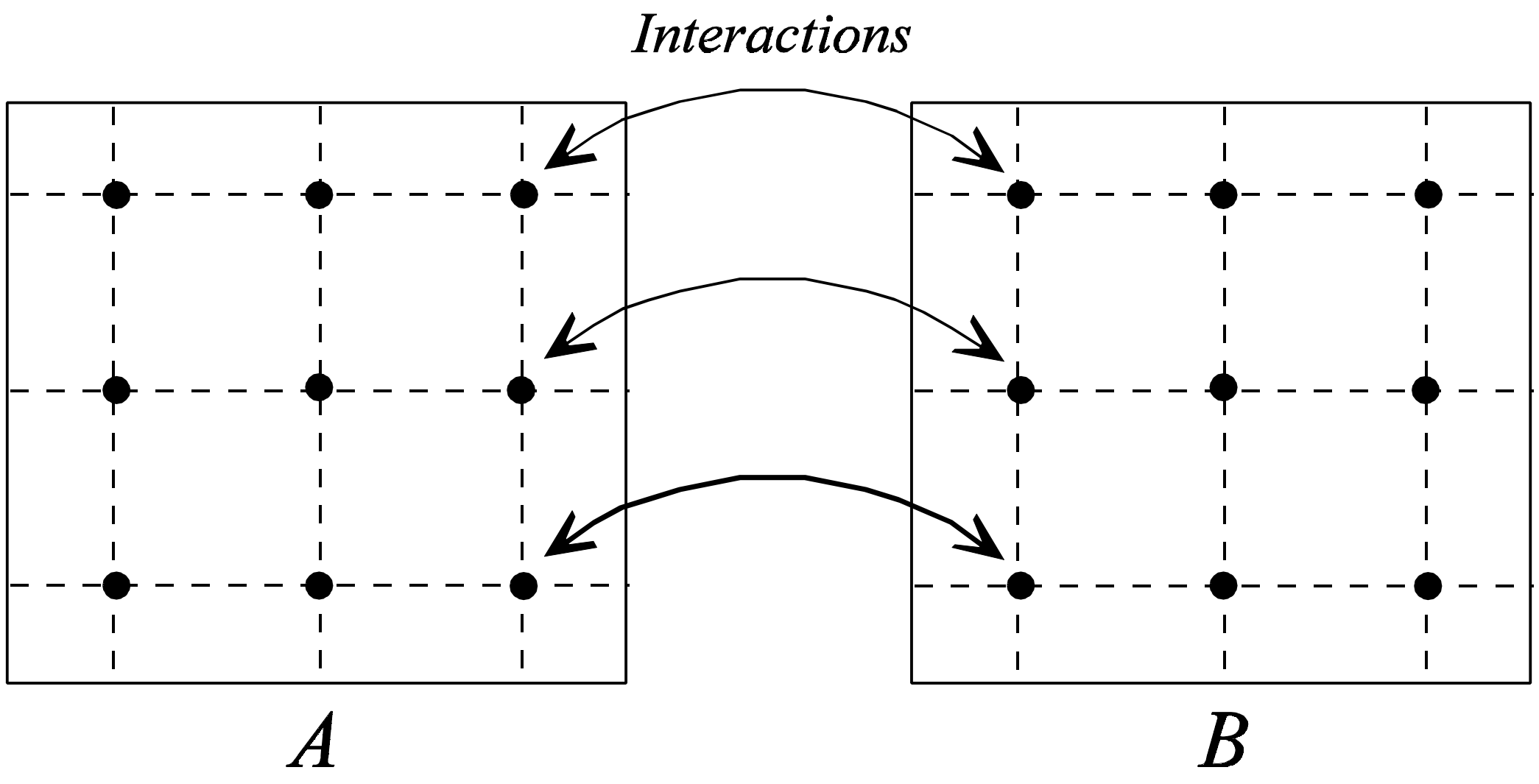

**Figure 5** - When two systems combine, cross interactions of the particles, especially those near the joining region, reduce the internal potential energy of the combined system.